\documentclass[prb,aps,twocolumn,showpacs,nofootinbib]{revtex4}

\usepackage{graphicx}
\usepackage{stmaryrd}
\usepackage{rotating}
\usepackage{amsmath}
\usepackage{amsfonts}
\usepackage{amssymb}
\usepackage[countmax]{subfloat}
\usepackage{dcolumn} % align table columns on decimal point
\usepackage{bm} % bold math
\usepackage{color}

\usepackage{float}
\usepackage{latexsym}

\begin{document}

\title{Scaling Theory of Topological Phase Transitions}

%\title{Scaling Theory of Berry Curvature}
%\title{Judging Topology from Berry Curvature Renormalization}

\author{Wei Chen} 

\affiliation{Institut f\"ur Theoretische Physik, ETH Z\"urich, 8093 Z\"urich, Switzerland}

\date{\rm\today}

\begin{abstract}

{Topologically ordered systems are characterized by topological invariants that are often calculated from the momentum space integration of a certain function that represents the curvature of the many-body state. The curvature function may be Berry curvature, Berry connection, or other quantities depending on the system. Akin to stretching a messy string to reveal the number of knots it contains, a scaling procedure is proposed for the curvature function in inversion symmetric systems, from which the topological phase transition can be identified from the flow of the driving energy parameters that control the topology (hopping, chemical potential, etc.) under scaling. At an infinitesimal operation, one obtains the renormalization group (RG) equations for the driving energy parameters.  A length scale defined from the curvature function near the gap-closing momentum is suggested to characterize the scale invariance at critical points and fixed points, and displays a universal critical behavior in a variety of systems examined. }

%Within the same topological phase, the curvature function changes as external parameters vary, but its integration remains unchanged.

%Akin to stretching a messy string to reveal the number of knots it contains, a scaling procedure for the Abelian Berry curvature in topological insulators and superconductors with inversion symmetry is proposed, from which the topological phase transition can be identified from the flow of the energy parameters that control the topology (hopping, chemical potential, etc.) under scaling. The scaling procedure is valid for any dimension and symmetry class provided the topological invariant is calculated from the integration of Berry curvature. At an infinitesimal operation, one obtains the renormalization group (RG) equations for the driving energy parameters. A length scale defined from the Berry curvature near the gap closing points is suggested to characterize the scale invariance at critical points and fixed points, and displays a universal critical behavior near these points. 

\end{abstract}


%64.60.ae Renormalization-group theory

%64.60.F- Equilibrium properties near critical points, critical exponents

%73.20.-r Electron states at surfaces and interfaces

%74.90.+n: Other topics in superconductivity

\maketitle

\section{Introduction}

%{\cblue (1) Say the curvature function represents the curvature in $d$-dimensional manifold, whose precise form or how it is explicitly related to the wave function does not matter. }

In the great number of topological insulators (TI) and topological superconductors (TSC) discovered or proposed\cite{Haldane88,Kane05,Kane05_2,Bernevig06,Konig07,Read00,Qi09,Qi11,Hasan10}, the transition from topologically trivial to nontrivial phases can in general be triggered by varying all kinds of energy parameters, e.g., chemical potential\cite{Kitaev01,Lutchyn10,Oreg10,Mourik12}, hopping\cite{Su79}, interface coupling\cite{Choy11,NadjPerge13,Braunecker13,Pientka13,Klinovaja13,Vazifeh13,Rontynen14,Kim14,Sedlmayr15,Chen15}, etc. The canonical way of identifying the topological phase transition is to first consult the dimension and symmetry class\cite{Schnyder08,Kitaev09} of the system to see if topologically nontrivial phases are possible, and if so then seek for the value of the driving energy parameter at which the spectral gap closes at certain momenta, since topological phase transition necessarily involves band inversion. Ideally, the topological invariant jumps discretely when the system undergoes a topological phase transition, therefore it seems rather ambiguous to identify any asymptotic critical behavior from the topological invariant itself.

Drawing analogy from the Landau order parameter paradigm, a profound question is whether there exists a universal scaling scheme that can judge the topological phase transitions driven by any energy parameter in any dimension and symmetry class, and if the scaling procedure is by any means different from Kadanoff's scaling theory\cite{Kadanoff66} since the systems under question possess no Landau order parameter. Moreover, it would be of tremendous usage if the scaling scheme also renormalizes the driving energy parameter. The proposal related to entanglement entropy first shed a light on this issue\cite{Kitaev06,Levin06,Jiang12}. On the other hand, since a great number of TIs and TSCs are characterized by the topological invariants that are often calculated from the momentum space integration of a certain integrand function that represents the curvature of the many-body state, it is natural to suspect that a scaling scheme that renormalizes the integrand function can be used to judge topology. Depending on the system, the integrand function may be Berry curvature\cite{Berry84,Xiao10,Thouless82}, Berry connection\cite{Zak89}, or any other quantity whose integration gives the topological invariant. For the sake of a general discussion, hereafter the integrand function is referred to as the curvature function, and the topological invariant calculated from the integration of the curvature function is referred to as the winding number (to draw analogy to the number of knots in a closed string, and to distinguish it from other means of calculating topological invariants). If a scaling scheme for the curvature function exists and gives a RG flow of the driving energy parameter that judges the topological phase transition, the following criteria are expected: (1) The fixed point in the parameter space of the driving energy parameter has a particular configuration of curvature function that is invariant under the scaling procedure. (2) Any point in the parameter space flows into some fixed point while preserving the winding number during the flow. (3) Different fixed points correspond to different winding numbers.

%{\cblue (1) Check Murakami11 and see what he says about the universal phase diagram of TI. }

In this article, we propose a scaling procedure that satisfies the aforementioned criteria for inversion-symmetric systems. The motivation comes from a simple observation: To know the number of knots that a messy string contains, one can either integrate the curvature along the string (integrating the curvature function to get winding number), or simply stretch the string until the knots become obvious (the proposed scaling procedure). A single equation is proposed to implement this knot-tying principle in TI and TSC: Let $F({\bf k},\Gamma)$ be the curvature function at momentum ${\bf k}$ from which the winding number in $d$-dimension ${\cal C}=\int d^{d}{\bf k}\;F({\bf k},\Gamma)$ is calculated, with $\Gamma$ the driving energy parameter that controls the topology. Given a $\Gamma$ in the parameter space, we seek for a new $\Gamma^{\prime}$ that satisfies 
\begin{eqnarray}
F({\bf k}_{0},\Gamma^{\prime})=F({\bf k}_{0}+\delta{\bf k},\Gamma)
\label{scaling_scheme_general}
\end{eqnarray}
where ${\bf k}_{0}$ is a high symmetry point, and $\delta{\bf k}$ is a small deviation away from it satisfying $F({\bf k}_{0}+\delta{\bf k},\Gamma)=F({\bf k}_{0}-\delta{\bf k},\Gamma)$. As we shall see in the examples below, the operation of Eq.~(\ref{scaling_scheme_general}) in inversion-symmetric TI or TSC correctly captures the topological phase transition driven by any energy parameter. The fixed point is reached by iteratively solving $\Gamma$, and by considering a small deviation $\delta{\bf k}$, one obtains the RG equation for $\Gamma$ in the form of differential equation. The curvature function gradually evolves into a fixed point configuration that is invariant under this procedure, analogous to a string with its knots tight and cannot be stretched anymore. Moreover, for a great variety of models, an asymptotic universal critical behavior of the curvature function near the gap-closing momenta is revealed despite the system in general has no Landau order parameter or any short range correlation. The critical behavior may be varified by ultracold atoms in optical lattices in the case that the curvature function is the Berry curvature.

\section{Deviation-reduction mechanism}

To explain the mechanism behind Eq.~(\ref{scaling_scheme_general}) and demonstrate its analogy to knot-tying, consider a inversion-symmetric TI or TSC defined on a $d$-dimensional cubic lattice. If the system at $\Gamma$ and at the fixed point $\Gamma_{f}$ have the same topology, then the curvature function at $\Gamma$ can be expanded by 
\begin{eqnarray}
&&F({\bf k},\Gamma)=F_{f}({\bf k},\Gamma_{f})+F_{v}({\bf k},\Gamma)
\nonumber \\
&&=F_{f}({\bf k},\Gamma_{f})+\sum_{m_{1}}\sum_{m_{2}}...\sum_{m_{d}}\lambda_{{\bf m},\Gamma}\left(\prod_{i=1}^{d}\cos m_{i}k_{i}\right)\;,
\label{Berry_curvature_d_dimension}
\end{eqnarray}
where $F_{f}({\bf k},\Gamma_{f})$ is the fixed point curvature function that satisfies $F_{f}({\bf k}_{0},\Gamma_{f})=F_{f}({\bf k}_{0}+\delta{\bf k},\Gamma_{f})$, i.e., it is invariant under the operation of Eq.~(\ref{scaling_scheme_general}), and $F_{v}({\bf k},\Gamma)$ is the deviation away from this fixed point configuration at $\Gamma$, which has Fourier component $\lambda_{{\bf m},\Gamma}\equiv\lambda_{m_{1},m_{2}...m_{d},\Gamma}$. The expansion of $F_{v}({\bf k},\Gamma)$ is sound because it must not contribute to the integration of winding number 
\begin{eqnarray}
{\cal C}=\int d^{d}{\bf k}F({\bf k},\Gamma)=\int d^{d}{\bf k}F_{f}({\bf k},\Gamma_{f})\;,
\label{winding_number_integration}
\end{eqnarray} 
such that the system at $\Gamma$ and at $\Gamma_{f}$ have the same topology. Suppose we choose a particular high symmetry point such as ${\bf k}_{0}=(0,\pi,0,0,...\pi)$, and a small displacement away from it along $j$-th coordinate $\delta{\bf k}=(0,0,..\delta k_{j}..,0,0)$. Now we apply the operation of Eq.~(\ref{scaling_scheme_general}) by using Eq.~(\ref{Berry_curvature_d_dimension}). Expanding around ${\bf k}_{0}$ gives 
\begin{eqnarray}
&&F_{v}({\bf k}_{0},\Gamma^{\prime})-F_{v}({\bf k}_{0},\Gamma)
\nonumber \\
&&=-\frac{1}{2}m_{j}^{2}\delta k_{j}^{2}\sum_{m_{1}}\sum_{m_{2}}...\sum_{m_{d}}{\rm sgn}_{\bf m,k_{0}}\lambda_{{\bf m},\Gamma}\;,
\label{expansion_F}
\end{eqnarray}
where ${\rm sgn}_{\bf m,k_{0}}$ is the sign of $\prod_{i=1}^{d}\cos m_{i}k_{i}$ at the chosen ${\bf k}_{0}$ and at the Fourier component ${\bf m}=(m_{1},m_{2}...m_{d})$. Note that $F_{v}({\bf k}_{0},\Gamma)$ is the deviation at ${\bf k}_{0}$. Writting $F_{v}({\bf k}_{0},\Gamma^{\prime})-F_{v}({\bf k}_{0},\Gamma)=dF_{v}({\bf k}_{0},\Gamma)$ and introducing a pseudo scaling parameter $d{\tilde l}=dk_{j}$, one obtains 
\begin{eqnarray}
\frac{dF_{v}({\bf k}_{0},\Gamma)}{d{\tilde l}}=\left.\frac{1}{2}\partial_{k_{j}}F_{v}({\bf k},\Gamma)\right|_{{\bf k}={\bf k}_{0}+\delta{\bf k}}
\label{RG_equation_d_dimension}
\end{eqnarray}
This means the deviation at ${\bf k}_{0}$ changes under this operation according to the slope of the deviation at ${\bf k}_{0}+\delta{\bf k}$. If the initial value $\Gamma$ gives $F_{v}({\bf k}_{0},\Gamma) >0 \;(<0)$, i.e., a positive (negative) deviation at ${\bf k}_{0}$, then $F_{v}({\bf k},\Gamma)$ must curve down (curve up) as moving from ${\bf k}_{0}$ to ${\bf k}_{0}+\delta{\bf k}$ in order to conserve the winding number, so $\left.\partial_{k_{j}}F_{v}({\bf k},\Gamma)\right|_{{\bf k}={\bf k}_{0}+\delta{\bf k}}<0\;(>0)$ and consequently $|F_{v}({\bf k}_{0},\Gamma)|$ is reduced under this scaling procedure. Thus continuously applying Eq.~(\ref{scaling_scheme_general}) makes $F_{v}({\bf k},\Gamma)$ approaching zero and $F({\bf k},\Gamma)$ approaching $F_{f}({\bf k},\Gamma_{f})$. In other words, the principle behind Eq.~(\ref{scaling_scheme_general}) is that the deviation of the curvature function from its fixed point configuration is gradually reduced under this scaling procedure. As we shall see below, in $d=1$ systems, this is synonymous to stretching a messy string to reduce its messiness until the knots are tight. In general, There could be a situation that $F_{v}({\bf k}_{0},\Gamma) >0$ and $\left.\partial_{k_{j}}F_{v}({\bf k},\Gamma)\right|_{{\bf k}={\bf k}_{0}+\delta{\bf k}}>0$ at a given $\Gamma$ so $|F_{v}({\bf k}_{0},\Gamma)|$ increases at the beginning of the scaling process, but $|F_{v}({\bf k}_{0},\Gamma)|$ must eventually be reduced because the scaling process still leads to $\left.\partial_{k_{j}}F_{v}({\bf k},\Gamma)\right|_{{\bf k}={\bf k}_{0}+\delta{\bf k}}<0$ in order to conserve winding number.

Expanding Eq.~(\ref{scaling_scheme_general}) in both $d\Gamma=\Gamma^{\prime}-\Gamma$ and $\delta k_{j}^{2}\equiv dl$ yields the RG equation. In the leading order, 
\begin{eqnarray}
\frac{d\Gamma}{dl}=\frac{1}{2}\frac{\partial_{k_{j}}^{2}F({\bf k},\Gamma)|_{{\bf k}={\bf k}_{0}}}{\partial_{\Gamma}F({\bf k}_{0},\Gamma)}\;.
\label{generic_RG_equation}
\end{eqnarray}
This explains that the proper scaling parameter should be defined as quadratic in the displacement $\delta k_{j}^{2}\equiv dl$, whereas the $dk_{j}=d{\tilde l}$ in Eq.~(\ref{RG_equation_d_dimension}) is merely introduced to demonstrate the deviation-reduction mechanism. Since the RG equation involves only one variable $\Gamma$, it is always possible to map the RG flow into the motion of an overdamped particle in a conservative potential\cite{Chen04,Chang05}. Equation (\ref{Berry_curvature_d_dimension}), together with the fact that at critical point $\Gamma_{c}$ the curvature function diverges at the gap-closing momentum which is at a high symmetry point ${\bf k}_{0}$ if the system is inversion-symmetric\cite{Murakami11}, indicate that the curvature function near the gap-closing point ${\bf k}_{0}$ can be expanded by 
\begin{eqnarray}
F({\bf k}_{0}+\delta {\bf k},\Gamma)=\frac{F({\bf k}_{0},\Gamma)}{1\pm\xi^{2}|\delta {\bf k}|^{2}}
\label{correlation_length_definition}
\end{eqnarray}
As shown below, $\xi$ explicitly characterizes the scale invariance at the fixed point or critical point. Note that although Eq.~(\ref{correlation_length_definition}) is similar to the Ornstein-Zernike form of correlation function\cite{Ornstein14}, $\xi$ does not represent the correlation length of short range fluctuations which is obviously absent in TI and TSC. Instead, it is a length scale defined from the curvature function near the gap-closing point that signals the scale invariance under the proposed scaling scheme. Alternatively, one may use its inverse $\kappa=1/\xi$ as a momentum scale to characterize the scale invariance, which works equally well.

Two remarks are made before we move to concrete examples. Firstly, as in any RG procedure, this scaling scheme itself does not give a meaning to the fixed point, i.e., it does not tell us whether the fixed point is topologically trivial or nontrivial, which may nevertheless be clarified by direct calculation of Eq.~(\ref{winding_number_integration}) at $\Gamma_{f}$ or at any $\Gamma$ that flows to the $\Gamma_{f}$. Secondly, $\Gamma$ may increase or decrease along the RG flow, but the notion of relevant, irrelevant, or marginal coupling constants\cite{Shankar94} does not seem to apply to $\Gamma$, since the topological phase transitions are judged by the sign change of the RG equation in Eq.~(\ref{generic_RG_equation}) which is not directly related to how $\Gamma$ changes under the scaling procedure.

%One application of this scaling procedure is that it reduces the $d$-dimensional integration of winding number in Eq.~(\ref{winding_number_integration}) to a zero dimensional recursive operation of Eq.~(\ref{scaling_scheme_general}), which may greatly reduce the numerical expenses to judge topological order especially in higher dimensions. As we shall see in the examples below, in general one may need to apply Eq.~(\ref{scaling_scheme_general}) to multiple high symmetry points to capture multiple critical points in the $\Gamma$-space. Nevertheless, the cubic lattice we consider has $2^{d}$ high symmetry points in the first quartet, so one has only $2^{d}$ zero dimensional recursive equations that can scan through a large part of $\Gamma$-space with very little numerical effort.

\begin{figure}[ht]
\begin{center}
\includegraphics[clip=true,width=0.99\columnwidth]{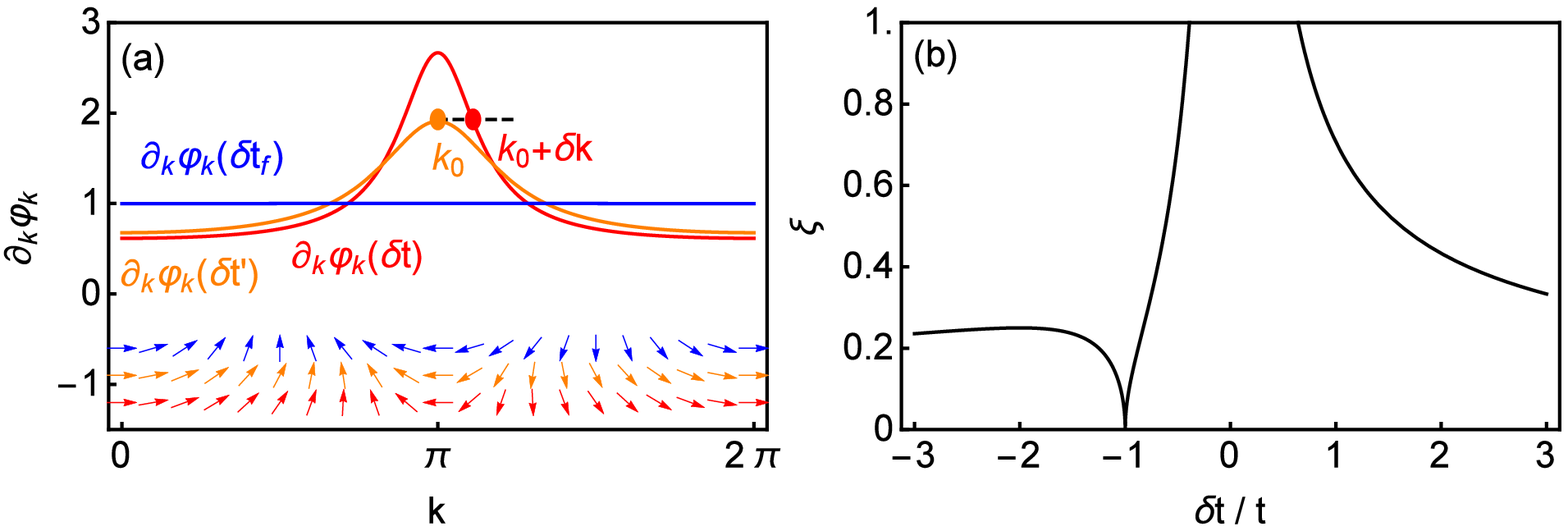}
\includegraphics[clip=true,width=0.99\columnwidth]{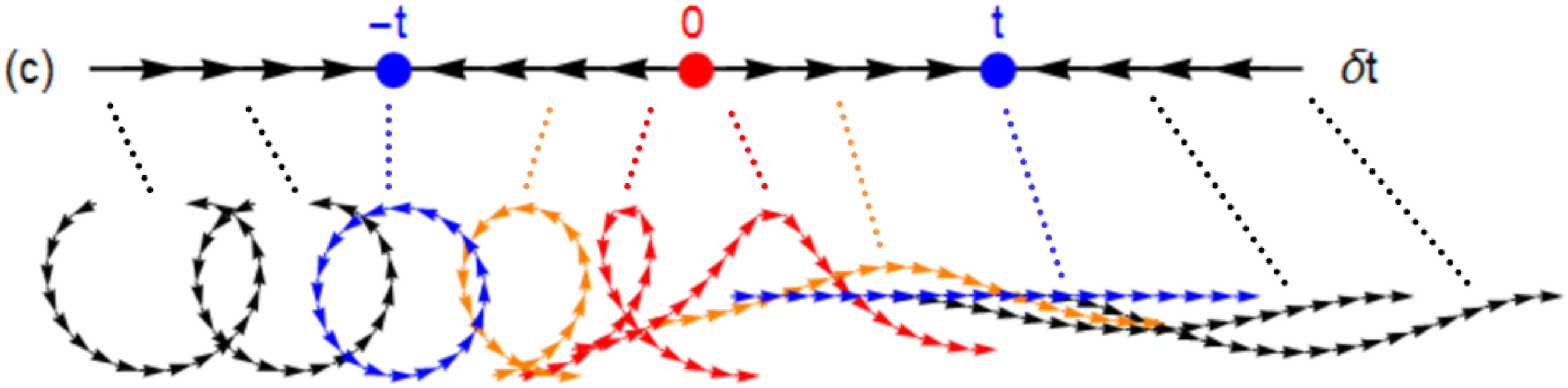}
\caption{ (a)The proposed scaling process, Eq.~(\ref{scaling_scheme_general}), applied to the topologically nontrivial phase of the SSH model with the choice $k_{0}=\pi$. Given an initial $\delta t$ and the corresponding Berry connection $\partial_{k}\varphi_{k}(\delta t)$ (red line), we find a new $\delta t^{\prime}$ by demanding $\partial_{k}\varphi_{k}(\delta t)$ at $k_{0}+\delta k$ to be equal to that of $\partial_{k}\varphi_{k}(\delta t^{\prime})$ (orange line) at $k_{0}$, as indicated by the dash line. This procedure reduces the deviation away from and leads to the fixed point configuration $\partial_{k}\varphi_{k}(\delta t_{f})$ (blue line). The $\varphi_{k}$ expressed as a vector field in the complex space for each configuration is indicated by colored arrows. (b) The length scale $\xi$ defined in Eq.~(\ref{correlation_length_definition}) as a function of $\delta t$. (c) The RG flow of $\delta t$. When joining the $\varphi_{k}$ arrows in (a) at a particular $\delta t$ head to tail in sequence to form a string, this scaling procedure resembles stretching the string to reveal whether it has a knot, as indicated by color arrows. } 
\label{fig:SSH_model_scaling}
\end{center}
\end{figure}

\section{Applications}

\subsection{Su-Schrieffer-Heeger (SSH) model}

To demonstrate the scaling scheme in $d=1$, we start from the spinless SSH model\cite{Su79} with periodic boundary condition (PBC), described by the Hamiltonian
\begin{eqnarray}
H&=&\sum_{i}(t+\delta t)c_{Ai}^{\dag}c_{Bi}+(t-\delta t)c_{Ai+1}^{\dag}c_{Bi}+h.c.
\label{SSH_Hamiltonian}
\end{eqnarray}
The topology of this model is judged by the winding number ${\cal C}$ of the operator $h_{k}(\delta t)=(t+\delta t)+(t-\delta t)e^{-ik}=|h_{k}(\delta t)|e^{-i\varphi_{k}(\delta t)}$ in the complex space when one goes through the entire Brillouin zone (BZ). Defining $q_{k}=h_{k}/|h_{k}|=e^{-i\varphi_{k}(\delta t)}$, the winding number is calculated from the curvature function which is the Berry connection\cite{Zak89} $\partial_{k}\varphi_{k}(\delta t)$ in this case, 
\begin{eqnarray}
{\cal C}&=&\frac{i}{2\pi}\oint dk \left(q_{k}^{-1}\partial_{k}q_{k}\right)=\frac{1}{2\pi}\oint dk \;\partial_{k}\varphi_{k}(\delta t)\;.
\label{Chern_number_SSH}
\end{eqnarray}
We now search for a new $\delta t'$ by applying Eq.~(\ref{scaling_scheme_general})
\begin{eqnarray}
\partial_{k}\varphi_{k}(\delta t')|_{k=k_{0}}=\partial_{k}\varphi_{k}(\delta t)|_{k=k_{0}+\delta k}
\label{Berry_curvature_scaling_SSH}
\end{eqnarray}
Writing $\delta t'-\delta t=d\delta t$, $(\delta k)^{2}=dl$, and using $\partial_{k}\varphi_{k}(\delta t)=\partial_{k}\arctan\left({\rm Im}h_{k}^{\ast}/{\rm Re}h_{k}^{\ast}\right)$ one obtains the RG equation at the leading order
\begin{eqnarray}
\frac{d\delta t}{dl}&=&\frac{\delta t}{4}\left(1-\frac{\delta t^{2}}{t^{2}}\right)
\;\;{\rm if}\;k_{0}=0\;,
\nonumber \\
\frac{d\delta t}{dl}&=&\frac{t^{2}}{4\delta t}\left(1-\frac{\delta t^{2}}{t^{2}}\right)
\;\;{\rm if}\;k_{0}=\pi\;,
\label{RG_eq_SSH}
\end{eqnarray}
both reproduce the correct critical points at $\delta t_{c}=0$ and fixed points at $\delta t_{f}=\pm t$, as indicated by the RG flow in Fig.~\ref{fig:SSH_model_scaling}. Through directly calculating the winding number, the $\delta t_{f}=t$ fixed point is topologically trivial and the $\delta t_{f}=-t$ fixed point is nontrivial. The example shown in Fig.~\ref{fig:SSH_model_scaling} (a) (red line) clearly demonstrates the deviation-reduction mechanism of Eq.~(\ref{RG_equation_d_dimension}), where the initial value $\Gamma=\delta t$ gives $F_{v}(k_{0},\Gamma)>0$ and $\partial_{k}F_{v}(k,\Gamma)|_{k=k_{0}+\delta k}<0$ such that $F_{v}(k_{0},\Gamma)$ is gradually reduced to zero under the operation of Eq.~(\ref{scaling_scheme_general}). The choice of different $k_{0}$ means different $F_{v}(k_{0},\Gamma)$ to start with, hence the speed of converging to the fixed point configuration is also different, as reflected in the two equations in  Eq.~(\ref{RG_eq_SSH}). Note that the phase transition occurs when the gap at $k_{0}=\pi$ closes, but the scaling scheme works in this model even if one chooses $k_{0}=0$ that is not the gap-closing point.

Alternatively, one can introduce the ratio $\gamma=(t-\delta t)/(t+\delta t)$ and discuss the topological phase transition upon tuning $\gamma$. Using Eq.~(\ref{scaling_scheme_general}) with $k_{0}=0$ leads to 
\begin{eqnarray}
\frac{d\gamma}{dl}=\beta(\gamma)=\frac{\gamma}{2}\left(\frac{\gamma-1}{\gamma+1}\right)
\end{eqnarray}
which well reproduces the two critical points at $\gamma=\pm 1$. Thus the proposed scaling scheme is valid whether the system is parametrized by $\delta t$ or $\gamma$.

%A Ginzburg-Landau (GL) type of description can be constructed for the Chern number near the critical points $\gamma=\pm 1$. Writing the RG equation as $d\gamma/dl=\beta(\gamma)$ and introducing 
%\begin{eqnarray}
%f=\left(\frac{\beta(\gamma)}{|\beta(\gamma)|}-1\right){\cal C}+{\cal C}^{2}
%\end{eqnarray}
%one can use $\partial f/\partial {\cal C}=0$ to obtain
%\begin{eqnarray}
%{\cal C}=\frac{1}{2}\left(\frac{\beta(\gamma)}{|\beta(\gamma)|}-1\right)
%\end{eqnarray}
%which is the correct Chern number near the critical points. {\cblue(There must be a $\pm$ difference between $\gamma=1$ and $\gamma=-1$, which I ignore at this moment.)}

%{\cblue (1) I realized that this $\cos mk$ expansion only works for lattice model. For the continuum model, as I addressed later for Chern insulator and topological SC, there's only one high symmetry point $k=0$ and the upper bound of integration is at $k=\infty$. Then how should I expand a function like this?  }

%{\cblue (1) I realized that SSH model is also defined on a lattice and has two high symmetry points. So it seems like one should do RG on both high symmetry points. Or just comment that doing RG on both points give the same result. Check this. }

From Eq.~(\ref{correlation_length_definition}) we obtain the $\xi$ deduced from the gap-closing point $k_{0}=\pi$
\begin{eqnarray}
\xi&=&\left|\frac{t}{4\delta t}\left(1+\frac{t}{\delta t}\right)\right|^{1/2}\;,
\end{eqnarray}
As shown in Fig.~\ref{fig:SSH_model_scaling}, $\xi=\infty$ at the critical point $\delta t=0$, and $\xi=0$ at the topologically nontrivial fixed point $\delta t=-t$, both signature the scale invariance\cite{Huang87}. We emphasize that the scaling scheme here is a procedure akin to knot-tying, and scale-invariance means that the curvature function converges to a configuration analogous to a string with all its knots tight (curvature function at gap-closing point stops changing), as indicated in Fig.~\ref{fig:SSH_model_scaling} (c), both have different meaning than those in Kadanoff's scaling theory. The topologically trivial fixed point $\delta t=t$ has $\xi=1/\sqrt{2}$, meaning that while the amplitude of $\partial_{k}\varphi_{k}(\delta t)$ is approaching zero everywhere, its functional form is approaching the first harmonic $\cos k$.

\begin{figure}[ht]
\begin{center}
\includegraphics[clip=true,width=0.99\columnwidth]{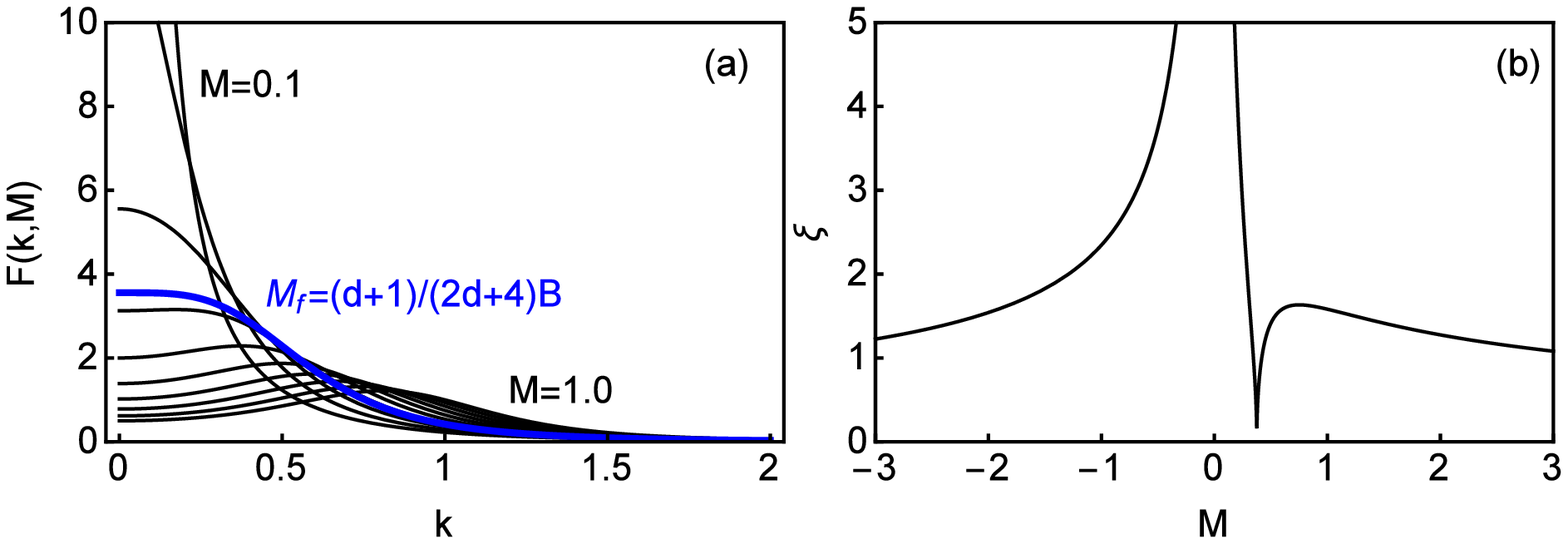}
\includegraphics[clip=true,width=0.9\columnwidth]{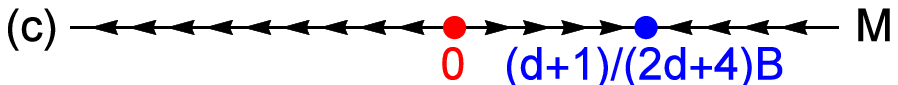}
\caption{(color online) (a) Berry curvature of the topologically nontrivial phase of $d=2$ Chern insulators in the continuum, which has the form of Eq.~(\ref{Chern_Berry_generic_form}), at evenly spaced values of $0.1\leq M\leq 1.0$. All the lines evolve to the fixed point configuration labeled by the blue line under the proposed scaling procedure. (b) The length scale $\xi$ defined in Eq.~(\ref{correlation_length_definition}) diverges at $M_{c}=0$ and vanishes at $M_{f}=(d+1)/(2d+4)B$. We set $B=1$ in these plots. (c) The RG flow of $M$.  } 
\label{fig:Chern_continuum_scalin}
\end{center}
\end{figure}

\subsection{Chern insulators in a continuum}

To demonstrate the feasibility of Eq.~(\ref{scaling_scheme_general}) in higher dimensions, we consider spinless Chern insulators in the $d$-dimensional continuum\cite{Bernevig13}, the low energy sector of which has the generic Dirac form
\begin{eqnarray}
H({\bf k})=\sum_{i}d_{i}({\bf k})\Gamma_{i}
\label{Dirac_Hamiltonian}
\end{eqnarray}
where $\Gamma_{i}$ satisfy the Clifford algebra. The curvature function from which the winding number is calculated, take $d=2$ and $d=4$ as examples, is
\begin{eqnarray}
&&F\propto\epsilon^{abc}{\hat d}_{a}\partial_{x}{\hat d}_{b}\partial_{y}{\hat d}_{c}\;,\;\;\;{\rm for}\;d=2
\nonumber \\
&&F\propto\epsilon^{abcde}{\hat d}_{a}\partial_{x}{\hat d}_{b}\partial_{y}{\hat d}_{c}\partial_{z}{\hat d}_{d}\partial_{v}{\hat d}_{e}\;,\;\;\;{\rm for}\;d=4
\label{Chern_Berry_general_generic_form}
\end{eqnarray}
which has the generic form
\begin{eqnarray}
F({\bf k},M)=\frac{M+Bk^{2}}{\alpha\left[k^{2}+\left(M-Bk^{2}\right)^{2}\right]^{\frac{d+1}{2}}}
\label{Chern_Berry_generic_form}
\end{eqnarray}
where the prefactor $\alpha$ depends on symmetry and dimension but is unimportant for our argument. Using Eq. (\ref{scaling_scheme_general}) with the only high symmetry point ${\bf k}_{0}=(0,0)$ and $\delta{\bf k}=(\delta k_{x},\delta k_{y})$, writing the energy parameter to be renormalized as $M'-M=dM$ while keeping $B$ constant, and defining $|\delta{\bf k}|^{2}=dl$, the leading order RG equation is
\begin{eqnarray}
\frac{dM}{dl}=\left(\frac{d+1}{2d}\right)\frac{1}{M}-\left(\frac{d+2}{d}\right)B
\label{RG_eq_2D_Chern_continuum}
\end{eqnarray}
which has generically a critical point at $M_{c}=0$, and the two fixed points $M_{f}=(d+1)/(2d+4)B$ and $M_{f}=-\infty$ assuming $B>0$. The length scale calculated from Eq.~(\ref{correlation_length_definition})
\begin{eqnarray}
\xi=\left|\frac{d+1}{2M^{2}}-(d+2)\frac{B}{M}\right|^{1/2}
\label{correlation_length_2D_Chern_continuum}
\end{eqnarray}
vanishes at the fixed points and diverges at the critical point. The results for $d=2$ spinless Chern insulators, which have $d_{1}=k_{x}$, $d_{2}=k_{y}$, and $d_{3}=M-Bk^{2}$ are shown in Fig.~\ref{fig:Chern_continuum_scalin}, where the curvature function is the Berry curvature\cite{Bernevig13}. In contrast to the knot-tying picture in $d=1$, the scaling procedure in $d=2$ is a process to stretch the skyrmion texture of ${\bf d}({\bf k})$ in Eq.~(\ref{Dirac_Hamiltonian}) without changing the skyrmion number\cite{Bernevig13} until the curvature function flattens to second order at the gap-closing point (blue line in Fig.~\ref{fig:Chern_continuum_scalin}(a)).

Two systems of similar kind are Haldane's $d=2$ graphene model\cite{Haldane88} and Kane-Mele model\cite{Kane05,Kane05_2}. Consider the spinless Haldane model whose expansion around ${\bf K}$ and ${\bf K}^{\prime}$ points of the reciprocal space of the hexagonal lattice is described by the Hamiltonian\cite{Bernevig13}
\begin{eqnarray}
h({\bf k}_{0}+\delta{\bf k})&=&-3t_{2}\cos\phi\pm\frac{3}{2}t_{1}\left(\delta k_{y}\sigma_{x}\mp\delta k_{x}\sigma_{y}\right)
\nonumber \\
&&+\left(M\mp 3\sqrt{3}t_{2}\sin\phi\right)\sigma_{z}
\end{eqnarray}
where upper sign is for ${\bf k}_{0}={\bf K}$ and the lower sign ${\bf k}_{0}={\bf K}^{\prime}$. Applying Eqs.~(\ref{Chern_Berry_general_generic_form}) and (\ref{scaling_scheme_general}) yields 
\begin{eqnarray}
\frac{dM}{dl}&=&\frac{3}{4}\frac{\left(3t_{1}/2\right)^{2}}{M\mp 3\sqrt{3}t_{2}\sin\phi}\;,
\nonumber \\
\xi&=&\left|\frac{3\left(3t_{1}/2\right)^{2}}{2\left(M\mp 3\sqrt{3}t_{2}\sin\phi\right)^{2}}\right|^{1/2}\;,
\end{eqnarray}
correctly reproducing the critical points at $M_{c}=\pm 3\sqrt{3}t_{2}\sin\phi$.

%{\cblue (1) For topological SC, the Berry curvature is defined differently. Use Schnyder's $(q\partial_{i} q)(q\partial_{j} q)(q\partial_{j} q)$. Maybe this chiral $p$-wave also includes Kitaev's 1D model, Check Bernevig book. }

\begin{figure}[ht]
\begin{center}
\includegraphics[clip=true,width=0.9\columnwidth]{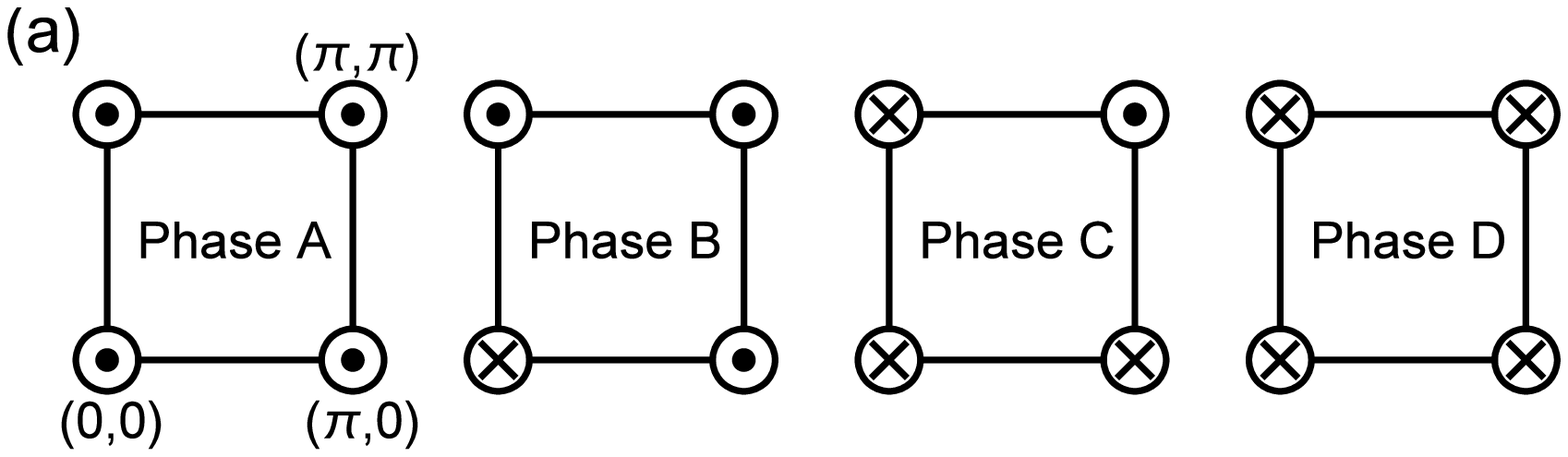}
\includegraphics[clip=true,width=0.9\columnwidth]{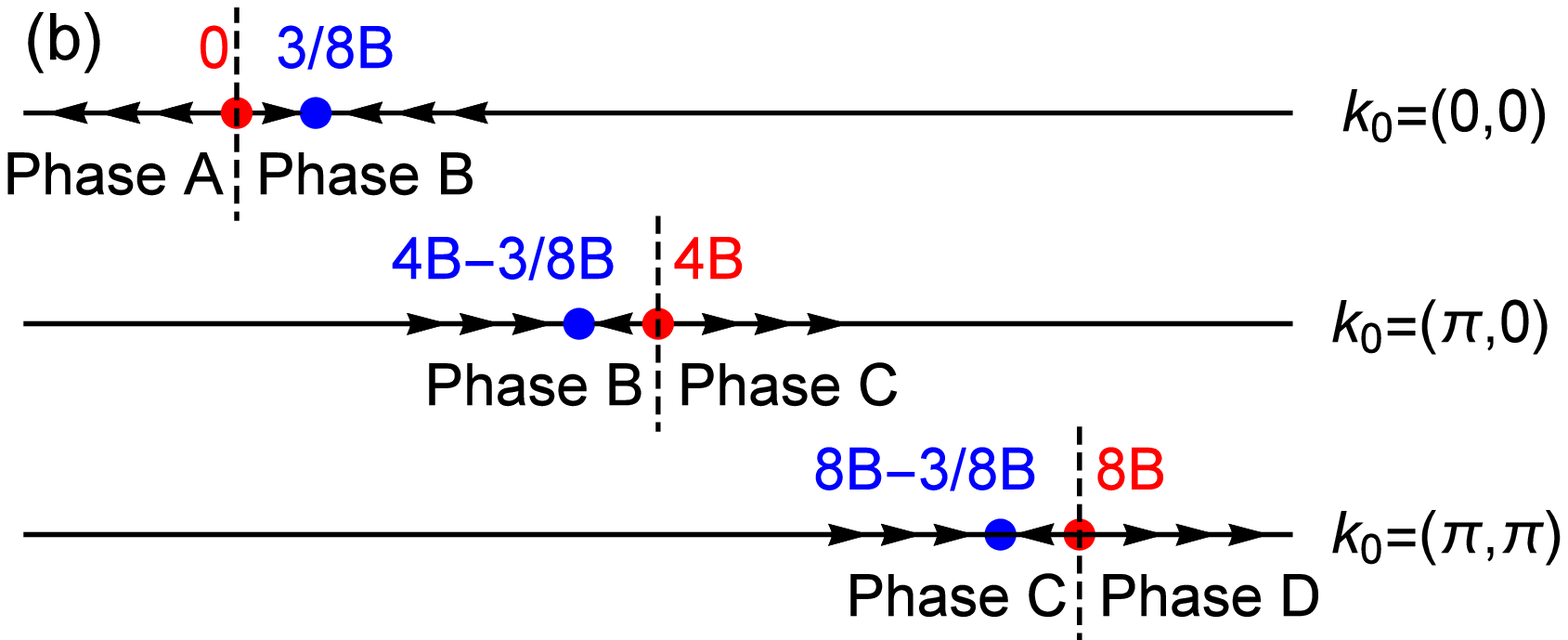}
\caption{ (a) The signs of $d_{3}({\bf k}_{0})$ at high symmetry points for the four phases of $d=2$ Chern insulators. (b) The RG flow of $M$ by choosing three different high symmetry points ${\bf k}_{0}$ with the same $\delta{\bf k}=(\delta k_{x},0)$. One sees that choosing a particular ${\bf k}_{0}$ captures the transition caused by gap-closing at this point, in accordance with the sign change of $d_{3}({\bf k}_{0})$ in (a).  } 
\label{fig:2D_Chern_d3}
\end{center}
\end{figure}

\subsection{Chern insulators on a cubic lattice}

To discuss the Chern insulators on a $d$-dimensional cubic lattice, we consider Eq.~(\ref{Dirac_Hamiltonian}) in $d=2$ and $d=4$ with components\cite{Bernevig13}
\begin{eqnarray}
{\bf d}=\left(\sin k_{x},\sin k_{y},....\sin k_{d},M-2B(d-\sum_{i=1}^{d}\cos k_{i})\right)\;.
\nonumber \\
\end{eqnarray}
The $2^{d}$ high symmetry points in the first quartet ${\bf k}_{0}=(k_{0x},k_{0y},...,k_{0d})$ consist of each $k_{0i}$ being either $0$ or $\pi$. The curvature function expanded around each ${\bf k}_{0}$ takes the form 
\begin{eqnarray}
F({\bf k}_{0}+\delta{\bf k},M)&=&\frac{(-1)^{N_{\pi}}\left[\left(M-4BN_{\pi}\right)+\Xi(\delta{\bf k})\right]}{\left\{|\delta{\bf k}|^{2}+\left[\left(M-4BN_{\pi}\right)-\Xi(\delta{\bf k})\right]^{2}\right\}^{\frac{d+1}{2}}}\;,
\nonumber \\
\Xi(\delta{\bf k})&=&B\left(\sum_{i\in 0}\delta k_{i}^{2}-\sum_{i\in \pi}\delta k_{i}^{2}\right)
\end{eqnarray}
where $\sum_{i\in 0}$ ($\sum_{i\in \pi}$) denotes summation over $\delta k_{i}$ at which $k_{0i}=0$ ($k_{0i}=\pi$), and $N_{\pi}$ is the number of $\pi$'s in $\left\{k_{0i}\right\}$. Applying Eq.~(\ref{scaling_scheme_general}) with $\delta{\bf k}=\delta k_{s}{\hat{\bf s}}$ along one particular coordinate ${\hat{\bf s}}$ yields 
\begin{eqnarray}
\frac{dM}{dl}=\left(\frac{d+1}{2d}\right)\frac{1}{M-4BN_{\pi}}\mp\left(\frac{d+2}{d}\right)B\;,
\end{eqnarray}
and hence the generic fixed points $M_{f}=4BN_{\pi}\pm (d+1)/(2d+4)B$ and $M_{f}=\mp\infty$ when choosing a particular ${\bf k}_{0}$. The length scale $\xi$ defined from the gap-closing point satisfies 
\begin{eqnarray}
\xi=\left|\left(\frac{d+1}{2}\right)\frac{1}{\left(M-4BN_{\pi}\right)^{2}}\mp\frac{(d+2)B}{M-4BN_{\pi}}\right|^{1/2}
\end{eqnarray}
The top sign in each $\pm$ or $\mp$ corresponds to the case when the component of ${\bf k}_{0}$ in the scaling direction $\delta{\bf k}=\delta k_{s}{\hat{\bf s}}$ is $k_{0s}=0$, and the bottom sign is when $k_{0s}=\pi$. Because $N_{\pi}$ takes any integer value from $0$ to $d$, generically there are $d+1$ critical points located at $M_{c}=4BN_{\pi}$. The phase transition at a particular $M_{c}$ takes place when the gap closes\cite{Bernevig13,Murakami11} at one set of ${\bf k}_{0}$'s that have the same $N_{\pi}$, in accordance with applying Eq.~(\ref{scaling_scheme_general}) at these ${\bf k}_{0}$'s to capture the change of curvature function near them, as shown for $d=2$ in Fig.~\ref{fig:2D_Chern_d3}. The chiral $p$-wave superconductors in the continuum and in the lattice\cite{Schnyder08} have the same generic form of curvature function as the Chern insulators\cite{Bernevig13} (with the replacement $M\rightarrow$ chemical potential and $B\rightarrow 1/2\times$effective mass), and hence practically the same critical points and critical behavior. We remark that the SSH model\cite{Atala13} and Haldane's graphene model\cite{Jotzu14} have been realized by ultracold atoms in optical lattices, where the systems can be driven close to the critical point. The predicted critical behavior of $\xi$ can be verified in the systems where the curvature function is the Berry curvature, which may be measurable by detecting the anomalous velocity associated with the Berry curvature\cite{Jotzu14} or momentum space interferometry techniques\cite{Abanin13,Duca15}.

\section{Conclusions}

In summary, we present a scaling procedure to judge topological phase transitions driven by any energy parameter. The procedure is valid for inversion-symmetric models in any dimension and symmetry class provided the topological invariant is calculated from the integration of a certain curvature function. Our formalism reveals that the concept of scaling in topologically ordered systems falls into a completely different realm than Kadanoff's scaling theory in the Landau order parameter paradigm. Based on a simple knot-tying picture, the scaling procedure renormalizes the curvature function while keeping the winding number intact, through which the RG flow of the driving energy parameter $\Gamma$ is obtained, as well as the RG equation, Eq.~(\ref{generic_RG_equation}), under an infinitesimal operation. In essence, the scaling procedure uses the divergence of second derivative of the curvature function at the gap-closing momentum to find the critical point, and uses the flattening of the second derivative to find the fixed point, both achieved under a single operation of Eq.~(\ref{scaling_scheme_general}). A length scale defined from the Berry curvature near the gap-closing momentum shows an asymptotic universal critical behavior that diverges at the critical point in first power and vanishes at the nontrivial fixed point in square root if it is finite
\begin{eqnarray}
\xi\propto\left\{
\begin{array}{ll}
|\Gamma-\Gamma_{f}|^{1/2}|\Gamma-\Gamma_{c}|^{-1} & {\rm if}\;\Gamma_{f}\neq\pm\infty \\
|\Gamma-\Gamma_{c}|^{-1} & {\rm if}\;\Gamma_{f}=\pm\infty 
\end{array}
\right.
\end{eqnarray}
for a variety of inversion-symmetric systems examined. Applications to a broader class of models, such as those driven by interactions or inversion-asymmetric models, will be subject to future investigations.

%is shown to characterize the scale-invariance at the critical point $\Gamma_{c}$, whose critical behavior is found to be universal 
%\begin{eqnarray}
%\xi\propto\frac{1}{\left|\Gamma-\Gamma_{c}\right|}
%\end{eqnarray}

The author acknowledges the stimulating discussions with M. Sigrist, A. P. Schnyder, M. Aidelsburger, E. Demler, L. Glazman, and A. Stern.

\end{document}